\documentclass[pdflatex,sn-mathphys-num]{sn-jnl}

\usepackage{graphicx}%
\usepackage{multirow}%
\usepackage{amsmath,amssymb,amsfonts}%
\usepackage{amsthm}%
\usepackage{mathrsfs}%
\usepackage[title]{appendix}%
\usepackage{xcolor}%
\usepackage{textcomp}%
\usepackage{manyfoot}%
\usepackage{booktabs}%
\usepackage{algorithm}%
\usepackage{algorithmicx}%
\usepackage{algpseudocode}%
\usepackage{listings}%
\usepackage{dirtytalk}

\newcommand{\com}[2]{\left[#1,#2\right]}

\newcommand{\bla}{bla\\bla\\bla\\bla\\bla}

\newcommand{\mc}[1]{\mathcal{#1}}

\theoremstyle{thmstyleone}%
\theoremstyle{thmstyletwo}%

\theoremstyle{thmstylethree}%

\begin{document}


\title[Article Title]{Over forty years of research towards the understanding of Quantum Brownian Motion -- the contributions of A. O. Caldeira}


\author*[1]{\fnm{Marcus V. S.} \sur{Bonan\c{c}a}}\email{mbonanca@ifi.unicamp.br}

\author[2,3,4]{\fnm{Sebastian} \sur{Deffner}}\email{deffner@umbc.edu}

\author[5]{\fnm{Gert-Ludwig} \sur{Ingold}}\email{gert.ingold@physik.uni-augsburg.de}

\affil*[1]{\orgdiv{Instituto de F\'isica Gleb Wataghin}, \orgname{Universidade Estadual de Campinas}, \orgaddress{\city{Campinas}, \state{S\~{a}o Paulo}, \postcode{13083-859}, \country{Brazil}}}

\affil[2]{\orgdiv{Department of Physics}, \orgname{University of Maryland, Baltimore County}, \orgaddress{\city{Baltimore}, \state{Maryland}, \postcode{21250},  \country{USA}}}

\affil[3]{\orgdiv{Quantum Science Institute}, \orgname{University of Maryland, Baltimore County}, \orgaddress{\city{Baltimore}, \state{Maryland}, \postcode{21250},  \country{USA}}}

\affil[4]{\orgdiv{National Quantum Laboratory}, \orgaddress{\city{College Park}, \postcode{20740}, \state{Maryland}, \country{USA}}}

\affil[5]{\orgdiv{Institut für Physik}, \orgname{Universität Augsburg}, \postcode{86135} \orgaddress{\city{Augsburg},  \country{Germany}}}


\abstract{This article presents a brief account of Amir O. Caldeira's contributions to the theory of quantum Brownian motion. Motivated by its importance, we outline the description of Brownian motion in the quantum regime following Caldeira's first works. In this context, we particularly highlight the effect of dissipation on the tunneling rate out of a metastable state. We then journey along the alternative ways to approach quantum Brownian motion developed by Caldeira during his career, which go beyond the so-called Caldeira-Leggett model. We conclude by summarizing some of Caldeira's contributions to contemporary fields such as the theory of quantum decoherence and quantum thermodynamics, that were strongly inspired by his eponymous approach to quantum Brownian motion.}




\maketitle

\section{Introduction}

The observation of the seemingly random motion of small particles immersed in a liquid by Brown in 1827 \cite{Brown1828} and possibly others \cite{Ingenhousz1784} before him initially did not catch much attention of physicists. It took until 1888 that Gouy \cite{Gouy1888} reminded the physics community of this phenomenon and claimed that the particles' dynamics is caused by molecular motion in the liquid, although he thought that a partially coordinated motion of molecules is required. When Einstein published a study of the influence of molecular motion in a liquid  at finite temperature on an immersed particle in one of his famous papers of 1905 \cite{Einstein1905}, he stated in the introduction that this phenomenon might be identical to the Brownian motion, but he was not certain. 

An important result of the work of Einstein \cite{Einstein1905} and Sutherland \cite{Sutherland1905} is the relation
\begin{equation}
    D = \frac{k_\text{B}T}{6\pi\eta a}
    \label{eq:einsteinrelation}
\end{equation}
between the diffusion coefficient $D$, the temperature $T$, the friction coefficient $\eta$ and the radius $a$ of the spherical particle immersed in a liquid or gas, and $k_\text{B}$ is the Boltzmann constant. A microscopic derivation based on the collision of molecules with the spherical particle was presented not much later by Smoluchowski \cite{Smoluchowski1906}. The relation (\ref{eq:einsteinrelation}) represents a special case of the fluctuation-dissipation theorem, one of the cornerstones of non-equilibrium Statistical Mechanics \cite{Kubo1966}. Brownian motion is actually the paradigmatic example of irreversible phenomena, due to its simple modeling of the interplay between relaxation and thermal fluctuations. It thus plays a central role in the understanding and modeling of non-equilibrium processes, in general.    

The absence of Planck's constant in Eq.~(\ref{eq:einsteinrelation}) emphasizes that it is a result of classical physics. It should therefore not come as a surprise that the diffusion constant vanishes in the limit of zero temperature. In the quantum realm, however, the relevance of Brownian motion is attested by its manifestations in different physical phenomena, where quantum noise plays a pertinent role. In the field of quantum optics, for example, the decoherence of the quantum state of an atom coupled to a thermal radiation field is a central problem \cite{Zoller2004Book}. The quantum decay of metastable states in the presence of dissipation and genuine quantum noise plays a central role in chemical kinetics, diffusion in solids, and electron transport in semiconductors, to name just a few examples \cite{Hanggi1986JStatPhys}. In condensed matter and cold atom physics, the mobility of different types of impurities \cite{Giamarchi2012PRA,Xu2025PRL} or topological defects such as skyrmions \cite{Troncoso2014AnnPhys,Mochisuzi2014NatMat,Petrovic2025RMP} or magnetic domain walls has great relevance. In the last case, Brownian motion plays an important role in defect driven phase transitions \cite{Kosterlitz2017RMP}. 

Even more pertinently, the Nobel Prize in Physics 2025 was awarded for the discovery of macroscopic quantum tunneling, which in quantum circuits formed by superconducting junctions is also subject to dissipation \cite{Devoret1985}. The experimental verification of the dissipative effects on the tunneling rate by the 2025 Nobel Prize winners is particularly relevant in the present context, since it is exactly this quantum tunneling that motivated Caldeira's very first contribution to Brownian motion in the quantum regime. In a series of papers resulting from his doctoral thesis \cite{Caldeira1981PRL,Caldeira1982PRL,Caldeira1983}, Caldeira and his advisor Anthony J. Leggett developed  an approach to predict the effects of dissipation caused by a reservoir on the tunneling rate of a ``quantum particle". This approach was first applied to the specific problem of quantum circuits, but was soon recognized by the authors in Ref.~\cite{Caldeira1983PhysicaA} as a description of quantum Brownian motion in general. Their path-integral formulation of the problem allowed for a description from weak to strong coupling of the striking differences between quantum dynamics and its corresponding classical limit. For more details on the advantageous use of the path-integral scheme and a summary on subtle issues related to quantum Brownian motion, see Ref.~\cite{Hanggi2005Chaos}.

\section{Quantum Brownian Motion: system plus environment}
\label{sec:qbm}

While originally, the term ``Brownian motion'' was coined in the context of a particle immersed in a liquid or in a gas volume, the term is now understood in a more general sense where a physical degree of freedom interacts with an environment. In their 1983 paper discussing the path-integral approach to quantum Brownian motion, Caldeira and Leggett \cite{Caldeira1983PhysicaA} are motivated by an electrical circuit consisting of an inductance $L$, a capacitance $C$, and a resistance $R$, thereby raising several interesting aspects.

First of all, the system is no longer a small particle but the charge $Q$ on the capacitor, i.e. a collective degree of freedom related to the electron density in the circuit. As the equation for the charge dynamics in an $LCR$ circuit
\begin{equation}
    L\ddot Q + R\dot Q+\frac{Q}{C} = 0
    \label{eq:lcr_circuit}
\end{equation}
shows, this degree of freedom effectively moves in a harmonic potential. Here, we encounter an important difference to the original concept of Brownian motion where the system degree of freedom can be seen as a free damped particle. This difference is crucial when we take the step from classical to quantum Brownian motion. In the case of the free damped particle, only two frequency scales are present. The first one, $k_\text{B}T/\hbar$, arises from the environmental noise and is  related to the temperature. The second one is given by the friction coefficient. The transition into the quantum regime requires at least a third frequency, which could be arising from a confining volume or more generally some potential or for example from a memory time of the environment. Therefore, quantum Brownian motion requires a more general setting as the classical Brownian motion in the original sense.

Secondly, the environment is now represented by a resistor. It would be far too complicated and in fact unnecessary to employ a model for the true microscopic dynamics in the resistor. Caldeira and Leggett argue that it is sufficient to start from a Hamiltonian for system and environment which for a reduced description of the system degree of freedom alone yields a Langevin equation of the type
\begin{equation}
    m\ddot x + \eta \dot x +v'(x) = F(t)
    \label{eq:langevinequation}
\end{equation}
reproduced from Eq.~(1.1) of Ref.~\cite{Caldeira1983PhysicaA}. The force-force correlation function of the fluctuating force $F(t)$ needs to account for quantum fluctuations by respecting the fluctuation-dissipation theorem in its quantum version. On the left-hand side, $-v'(x)$ represents the force arising from a potential $v(x)$ in which a particle of mass $m$ moves.

In fact, this system-plus-environment approach became the standard working horse in Caldeira's research. This approach recognizes that, in general, no universal Langevin equation can be derived from a time-independent Hamiltonian or Lagrangian for the system degree of freedom alone. Rather, one has to choose a minimal model, coupling the system of interest to the thermal environment, such that classical Brownian motion is reproduced in the high-temperature limit. Thus, the choice of a system-plus-environment Hamiltonian appears quite logical in view of the microscopic understanding of Brownian motion. However, there were other approaches to the description of damped quantum systems like the use of a non-linear Schrödinger equation. Doubting either the correctness of the approach or their physical basis, Caldeira and Leggett opted for the system-plus-environment approach by choosing a Lagrangian of the form (cf. Eq.~(3.5) in \cite{Caldeira1983})
\begin{equation}
    L = \frac{1}{2}M\dot q^2 - V(q) + \frac{1}{2}\sum_j\left(m_j\dot x_j^2-m_j\omega_j^2x_j^2\right)-\sum_jF_j(q)x_j+\sum_j\frac{F_j^2(q)}{2m_j\omega_j^2}
    \label{eq:lagrangian}
\end{equation}
as the basis for their path-integral treatment. Here, $q$ denotes the system degree of freedom while $x_j$ refer to the environmental degrees of freedom. For a bilinear coupling, $F_j(q) = c_jq$.

With their choice, Caldeira and Leggett resorted to preceding work going back to Senitzky \cite{Senitzky1960PR,Senitzky1961PR}. In fact, there exists an even earlier publication by Magalinskiĭ \cite{Magalinskii1959JETP}, who considered a harmonic oscillator as system coupled bilinearly to a set of harmonic oscillators forming the environment and attributes the model to Bogolyubov, Terletskiĭ, and Toda. Relevant to us is that Magalinskiĭ by eliminating the environmental degrees of freedom derived a Langevin equation essentially of the form (\ref{eq:langevinequation}). We will come back to some differences between (\ref{eq:langevinequation}) and the result by Magalinskiĭ below.

In his study of dissipation in quantum mechanics, Senitzky \cite{Senitzky1960PR,Senitzky1961PR} had the radiation field of a lossy cavity in mind. Therefore, the system degree of freedom is represented by a harmonic oscillator. The environment is not specified, but it would have appeared as natural to model it again in terms of harmonic oscillators representing the external field modes. Ford, Kac, and Mazur in 1965 \cite{Ford1965JMP} considered a general system of harmonically coupled masses which by diagonalization can be related to the model already considered by Magalinskiĭ. A detailed analysis of a harmonic oscillator coupled bilinearly to a bath of harmonic oscillators as well as of a linear harmonically coupled chain was carried out by Ullersma in a series of four papers \cite{Ullersma1966Physica}. Finally in 1973, Zwanzig \cite{Zwanzig1973JStatPhys} allowed for a general system potential and made use of a counter-term in the Hamiltonian which deserves further discussion.

Coupling a system to an environment results in a damping term and a fluctuating force as we have seen in the Langevin equation (\ref{eq:langevinequation}). In addition, generally an additional term contributing to the potential will appear. This so-called potential renormalization is familiar as Lamb shift for atoms in a radiation field.
The potential renormalization can be avoided by including a corresponding term in the Hamiltonian (or Lagrangian) for system plus environment. This term is referred to as counter-term. The question therefore arises whether the counter-term should be included in the Hamiltonian or not.

This issue became highly relevant in the context of the question studied by Caldeira and Leggett \cite{Caldeira1981PRL}: How does dissipation influence the tunneling rate out of a metastable equilibrium at zero temperature? They found a reduction by a factor which depends exponentially on the damping strength, a result consistent with the expectation that dissipation renders the system more classical. In contrast, Widom and Clark \cite{Widom1982} found that dissipation will increase the tunneling rate. The reason for this apparent discrepancy lies precisely in the counter-term as explained in a comment by Caldeira and Leggett \cite{Caldeira1982PRL}. In short, Caldeira and Leggett on the one hand and Widom and Clark on the other hand have considered different situations. The latter did not take into account the counter-term in their Hamiltonian for system and environment. Therefore, the coupling between system and environment led to a renormalization of the bare potential resulting in a reduction of the barrier height. As a consequence, the tunneling rate will increase with increasing coupling. The situation considered by Caldeira and Leggett is different. Here, the potential for the system degree of freedom is the dressed potential and a counter-term thus needs to be included in the Hamiltonian. Then dissipation leads to a reduction of the tunneling rate as stated before.

The situation can be further exemplified by considering the free damped particle. In this case, the system degree of freedom in the presence of a coupling to the environment should not be subject to a potential. Therefore, the counter-term is needed in order to cancel the potential renormalization which would otherwise occur. That this is the correct approach to describe a free damped particle can be seen for the Lagrangian (\ref{eq:lagrangian}) by setting $F_j(q)=m_j\omega_j^2q$. The total Lagrangian can only be expressed in a translationally invariant way if the counter-term is included (see also Sec. 9 in \cite{Grabert1988PhysRep}). In order to correctly describe the experimental results by Devoret, Martinis, and Clarke on the decay of the zero-voltage state of a current-biased Josephson junction \cite{Devoret1985}, which led to the Nobel Prize in 2025, the counter-term needs to be included as well. Here, the washboard potential due to the Josephson junction is the one in presence of dissipation. In fact, the resistance appearing in (\ref{eq:lcr_circuit}) is an intrinsic physical property of the Josephson junction.

Given that the system-plus-environment Hamiltonian with the environment consisting of a set of harmonic oscillators coupled bilinearly to the system degree of freedom has had about two decades of history before it was used by Caldeira and Leggett, one may wonder why this model is nowadays routinely referred to as Caldeira-Leggett model (CLM). One of the reasons may be that the references cited above where well known in the quantum optics community but much less so in the condensed matter community. The motivation for the work by Caldeira and Leggett on quantum tunneling in dissipative systems \cite{Caldeira1981PRL,Caldeira1983} was the question whether quantum effects could manifest themselves for macroscopic systems \cite{Leggett1980SuppPTP,Leggett1984ContempPhys}. It was suggested to study macroscopic quantum tunneling and macroscopic quantum coherence of the phase of a superconducting condensate as system degree of freedom. In contrast to quantum optics where one often can employ a weak coupling approximation, this is typically not the case in condensed matter systems where the dissipative coupling can be strong. It is therefore not immediately obvious whether it is sufficient to describe the environmental degrees of freedom as harmonic oscillators bilinearly coupled to the environment. In appendix~C of Ref.~\cite{Caldeira1983}, Caldeira and Leggett argue that one needs to require a weak perturbation of the environmental degrees of freedom by the coupling to the system. They emphasize that this requirement is different from a requirement of weak damping of the system. A sufficiently large number of weakly perturbed environmental degrees of freedom can still lead to a strong effect on the system.

Another interesting aspect concerns the initial condition if one aims at studying the dynamics in quantum Brownian motion. In their study of quantum tunnelling in a dissipative system \cite{Caldeira1981,Caldeira1983}, Caldeira and Leggett could circumvent this issue because they determined the decay rate from the free energy. This quantity can be obtained from an imaginary-time path integral so that no real-time dynamics is involved. In the case of a metastable state, the free energy acquires a small imaginary part which determines the decay rate as established in previous work \cite{Langer1967,Callan1977}. The situation is different if one is interested in the real-time dynamics of a damped Brownian particle as in the study of the path integral approach to quantum Brownian motion by Caldeira and Leggett \cite{Caldeira1983PhysicaA}.

Typically, one is interested only in properties of the system degree of freedom and therefore traces over the environmental degrees of freedom. Within the path-integral formalism, such a procedure was first carried out by Feynman and Vernon \cite{Feynman1963}. The effect of the environment manifests itself then in an additional contribution to the action of the system degree of freedom, the so-called influence functional. This procedure becomes particularly simple by assuming factorizing initial conditions where initial correlations between system and environment are neglected. As a consequence, it is sufficient to evaluate an imaginary-time path integral for the environmental degrees of freedom while for the system degree of freedom, also two real-time path integrals need to be evaluated.

However, it is sometimes necessary to allow for initial correlations between system and environment. The dependence of the dynamics of the system degree of freedom on the initial conditions was first studied for the free Brownian particle \cite{Hakim1985PRA}. The basic idea is to start with an equilibrium state of system and environment and then to perform a measurement on the system degree of freedom. This could be a position measurement as proposed by Morais Smith and Caldeira \cite{Smith1987PRA,MoraisSmith1990} or a general measurement as was studied independently \cite{Grabert1988PhysRep}. Morais Smith and Caldeira also considered the possibility that the system degree of freedom is perturbed by an abrupt change in the system potential. Non-factorizing initial conditions allow for example for the evaluation of correlation functions. In the case of the damped harmonic oscillator, it became thus possible to give the quite complicated integral expressions found in \cite{Caldeira1983PhysicaA} a clear physical interpretation and thus a more transparent structure \cite{Grabert1988PhysRep}.

\section{Brownian motion in the quantum regime beyond the Caldeira-Leggett model}

As we have outlined in the preceding section, nowadays the Caldeira-Leggett model has become almost synonymous with the quantum description of Brownian motion. This is evident for instance when one checks how standard books on open quantum systems or quantum dissipative systems, such as Refs.~\cite{Breuer2007Book,Weiss2021Book}, approach the topic. This is of course strongly inspired by Caldeira and Leggett's seminal work \cite{Caldeira1983PhysicaA}. It is interesting to note, however, that Caldeira was from the beginning very critical about the range of applicability of the CLM to the different situations in which genuine quantum Brownian motion may take place. Quoting Caldeira and his collaborator Hedeg\r{a}rd in Ref.~\cite{Hedegard1987PhysScr}:

\say{\textit{...We fear that some of these authors overestimate the power of the model they investigate as far as possible applications are concerned. To give a specific example, we do not believe that a particle on a periodic potential, coupled to a bath of oscillators is a realistic model to describe the motion of charged particles in a solid. Let us try to clarify our point of view with a simple model. Suppose that we are interested in studying the diffusion of deep holes created in a X-ray absorption experiment. Once the hole is created, it interacts with the conduction electrons (the reservoir) and is allowed to hop to neighboring sites on the crystalline lattice. As far as X-ray absorption or emission spectra are concerned the treatment of this reservoir as an ensemble of fermions or oscillators (Tomonaga bosons) leads approximately to the same results. ...However, if we now allow the variable to reach values outside the above specified range, we no longer have the same bosons interacting with it; ...hopping over distances of the order of the lattice spacing or greater clearly makes the approximation of replacing the fermionic environment by a single set of oscillators inappropriate.}} Sec.~1, Page 610, left column.

The general picture of system-plus-environment, described by a global Hamiltonian in the quantum regime, was then understood by Caldeira as the situation in which quantum Brownian motion would always arise after the necessary trace (or average) over the reservoir's degrees of freedom, regardless of whether or not they are described by a set of harmonic oscillators. Indeed, this point of view is confirmed in some of Caldeira's subsequent publications such as Refs.~\cite{Hedegard1987PRB} and \cite{Barone1991PRA}. In particular, Ref.~\cite{Barone1991PRA} deals with the problem of electromagnetic radiation damping in the quantum but non-relativistic regime using the previously developed path-integral approach and the complete electron-plus-radiation-field Lagrangian. The authors explicitly mention in the introduction that this problem requires two important modifications of the previous treatment of quantum Brownian motion: a non-separable initial state of the total system as discussed at the end of Sec.~\ref{sec:qbm} and a spectral function of the bath arising from the microscopic model itself instead of being a phenomenological constraint as in \cite{Caldeira1983PhysicaA}. One of the important results reported in Ref.~\cite{Barone1991PRA} is the analysis of the decoherence of the eigenstates of a charge in an electromagnetic field. However, another interesting result is that in the classical limit the well-known Abraham-Lorentz equation, complete with a fluctuating force satisfying the fluctuation-dissipation theorem, is recovered, which confirms Caldeira's viewpoint about quantum Brownian motion as a phenomenon arising from general system-plus-environment setups.

Reference~\cite{Barone1991PRA} also highlights that the interaction between the electron and its radiation field is not described by a coordinate-coordinate potential energy, as in the CLM (see Eq.~(\ref{eq:lagrangian})). In contrast, when the electromagnetic field is converted into a set of modes and the description is changed to a Hamiltonian one, this interaction becomes bilinear in the coordinates of the charge and the generalized momentum of the field and a counter-term appears naturally. Therefore, there is no counter-term in the initial total Lagrangian. Surprisingly, this counter-term exactly cancels another term that appears in the path-integral calculation of the influence functional, allowing for the treatment of the situation in which the charge is not confined by an external potential energy. 

\subsection{Quantum Brownian motion due to scattering}

This coordinate-momentum bilinear coupling was extensively explored by Caldeira in his subsequent work for several situations of interest in condensed matter physics. Perhaps, this can be considered Caldeira's most important contribution to quantum Brownian motion after his original path-integral approach using the CLM. This point of view is corroborated by Ref.~\cite{CastroNeto1991PRL}, also published in 1991, whose title is ``\emph{A new model of dissipation in quantum mechanics}". Despite the emphasis on quantum dissipation, the title could fairly be 
``\emph{A new model of quantum Brownian motion}", given the contents of the work. The proposed model has significant differences compared to CLM, such as:
\begin{itemize}
    \item The coupling between particle and the reservoir degrees of freedom is bilinear but through the particle's \emph{momentum} (recall the previous discussion about Ref.~\cite{Barone1991PRA}); in the absence of an external potential, the total Hamiltonian is inherently translational invariant, a property that is not shared with the standard CLM;
    \item  This coupling is also designed to describe a \emph{scattering} process (which is local in space), i.e., the particle is scattered by the reservoir excitations, whose number is conserved; this is in striking contrast to the CLM where the particle is always coupled to all harmonic oscillators (there is not the notion of harmonic modes closer than others) of the reservoir whose number of excitations is not conserved.
\end{itemize}

In mathematical terms, the model Hamiltonian reads \cite{CastroNeto1991PRL}
\begin{eqnarray}
 H = \frac{\left[ p - h(a_{k}^{\dagger}, a_{k})\right]^{2}}{2 m} + V(q) + \sum_{k} \hbar \omega_{k} a^{\dagger}_{k} a_{k}\,,
 \label{eq.ScattModel}
\end{eqnarray}
where $p$, $q$ and $m$ denote, respectively, the momentum, coordinate, and mass of the particle, $V(q)$ accounts for a possible external potential, $a^{\dagger}_{k}$ and $a_{k}$ stand for the creation and annihilation operators of the reservoir excitations, respectively, with energy $\hbar \omega_{k}$. The coupling is described by the function
\begin{equation}
    h(a_{k}^{\dagger}, a_{k}) = \frac{\hbar}{m}\sum_{k, k'} G_{k k'} a^{\dagger}_{k} a_{k'}\,,
\end{equation}
with $G_{k k'}^{*} = G_{k' k}$ and $G_{k' k} = -G_{k k'}$. Observe that the Hamiltonian (\ref{eq.ScattModel}) commutes with the number operator $N = \sum_{k} a^{\dagger}_{k} a_{k}$. Last but not least, $a^{\dagger}_{k}$ and $a_{k}$ are considered boson operators, but the authors say that their treatment could be easily extended to fermionic modes (this was indeed tackled by the authors in a later work, as we will outline shortly).

The effective quantum dynamics of the particle is then derived within the same path-integral formalism as in Ref.~\cite{Caldeira1983PhysicaA}. The functional integrals leading to the influence functional are exactly evaluated due to the quadratic form of the Hamiltonian (\ref{eq.ScattModel}). However, its exact expression makes the remaining functional integrals, those required to have a closed expression for the superpropagator, unfeasible. At this point, the only approximation of the whole calculation is made. The influence functional is expressed in terms of a matrix that obeys a matrix equation of Dyson's type, the solution of which is obtained in the usual Born approximation (see Ref.~\cite{CastroNeto1991PRL}). The authors then restrict the analysis to the ``free" case, $V(q)=0$, which is much easier to handle in this model than in the CLM \cite{Hakim1985PRA,Aslangul1985,Schramm1987,Grabert1988PhysRep}. The effective action that leads to the superpropagator has the usual real and imaginary kernels that describe the effects of the reservoir. Both are expressed in terms of the scattering function \cite{CastroNeto1991PRL},
\begin{equation}
    S(\omega,\omega') = 2\pi \sum_{k, k'}G^{2}_{k k'} \delta(\omega-\omega_{k}) \delta(\omega'-\omega_{k'})\,,
    \label{eq.ScattFunc}
\end{equation}
which is related to the scattering of the reservoir excitations between states of frequencies $\omega$ and $\omega'$. This function plays the same role as the well-known spectral function in the CLM through which the Ohmic, sub-Ohmic or super-Ohmic regimes are introduced \cite{Caldeira1983PhysicaA}. The authors take the following continuum limit for Eq.~(\ref{eq.ScattFunc}),
\begin{equation}
    S(\omega,\omega') = \alpha\,\omega\,\omega' \Theta(\omega_{c}-\omega)\Theta(\omega_{c}-\omega')\,,
\end{equation}
which introduces a phenomenological constant $\alpha$ and a sharp high-frequency cutoff at $\omega_{c}$ (which could be the Debye frequency, if the bosons were phonons) described by step functions $\Theta(x)$. This choice leads to a memoryless dissipative kernel and Ohmic dissipation (in the limit of an infinite cutoff frequency), i.e., damping proportional to the velocity like in Eq.~(\ref{eq:lcr_circuit}). \cite{Caldeira1983PhysicaA,Weiss2021Book}.

The damping constant obtained from this analysis qualitatively contrasts with that obtained in the quantum Brownian motion described by the CLM in the Ohmic regime. It has a temperature dependence varying as $T^{4}$ or $T$ for low and high temperatures, respectively. The authors did not aim at describing any particular physical phenomena in Ref.~\cite{CastroNeto1991PRL} although concrete physical systems were addressed in later publications, but they point out on the last page of the paper that the model captures the temperature dependence of the mobility of large acoustic polarons in one dimension.

\subsection{Brownian motion of quantum solitons in condensed matter physics}

Reference~\cite{CastroNeto1991PRL} paved the way for a series of applications in relevant situations in condensed matter physics. Among them, the articles \cite{CastroNeto1992PRB,CastroNeto1993PRE} may justify a few detailed remarks. In Ref.~\cite{CastroNeto1992PRB}, the authors specifically address the dynamics of polarons in one dimension. Both optical and acoustic polarons are treated within the same framework in the strong-coupling limit using the path-integral method developed in \cite{Caldeira1983PhysicaA}. The mobility and diffusion coefficients as functions of temperature are calculated without appealing to kinetic theory. However, it is important to notice the considerable part of the manuscript dedicated to showing how the second-quantized form of the Hamiltonians describing electrons interacting with either optical or acoustical phonons can give rise to quantum solitons. Additionally, it is shown that the solitons (representing the polarons) are scattered by physically distinct thermal phonons and hence perform Brownian motion. Interestingly, the effective model obtained describes a single particle scattered by excitations of a reservoir, very much like the model proposed in Ref.~\cite{CastroNeto1991PRL} and discussed in the previous section. An important difference is that the absorption or emission of phonons by the polaron (that is, processes that do not conserve the number of excitations) are also included. The authors then restrict their further analysis to the regime in which these processes are negligible.

In the later work \cite{CastroNeto1993PRE}, the authors focused on systematizing the Brownian motion of quantum solitons in one dimension. It is clearly stated in the introduction of Ref.~\cite{CastroNeto1993PRE} that the main goal is to connect, at that time, two apparently different areas, namely, the quantum theory of solitons, also interested in their transport properties, and the quantum theory of dissipative systems, referring to what had been developed during the 1980s by people as Caldeira himself. The authors then show that, when quantizing classical translational invariant field theories with solitons using the so-called collective-coordinate method \cite{Rajaraman1982Book}, the coordinate describing the center of the soliton becomes coupled to the other, infinitely many, degrees of freedom, that together make up the field. It turns out that the soliton can then be viewed as a particle that undergoes Brownian motion due to the coupling to the other modes of the field as already exemplified in the previous work \cite{CastroNeto1992PRB} on the polaron. The authors derive the total Hamiltonian and show that it is of the same form as the one found in Ref.~\cite{CastroNeto1992PRB}. The particle is thus coupled to the reservoir modes through its momentum and scattering processes conserving the number of excitations are the main mechanism leading to Brownian motion but terms breaking the number conservation are also obtained.

The collective-coordinate method was then combined with the path-integral approach, derived in the context of the CLM, but reshaped in Ref.~\cite{CastroNeto1991PRL} to treat the scattering model. This led to a new framework that Caldeira and his co-workers used to tackle a relevant class of problems. To name some of them, Refs.~\cite{CastroNeto1994PRB,Caldeira1995PRB} were motivated by interest in the motion of ions in superfluid $\,^{3}\mathrm{He}$, the mobility of particles in metals, and the diffusion of heavy particles in solids; The creep of vortices in type-II superconductors was studied in Ref.~\cite{Smith1996PRB}; The dynamics of skyrmions in quantum Hall ferromagnets was discussed in Ref.~\cite{Ferrer2000PRB}; and the mobility and diffusion of a variety of topological defects mostly in magnetic systems were also considered in Refs.~\cite{Desposito2000PRB,Ferrer2001PRB,Juricic2004PRL,Juricic2005PRB}. In all these situations, the mobility of the resulting Brownian motion is qualitatively different from that obtained via the CLM where it is always temperature independent.

\subsection{Variations around the Caldeira-Leggett model}

The scattering mechanism discussed previously was not the only variation explored by Caldeira and his co-workers. 
Predictions differing from those of the CLM were also obtained from certain modifications considered in Refs.~\cite{Caldeira1993PRB,Duarte2006PRL}. 

In Ref.~\cite{Caldeira1993PRB}, the reservoir is modeled by a set of independent two-level systems that are individually coupled to the particle through a bilinear potential energy depending on the particle's coordinate and their Pauli operators $\sigma_{x}$. The path-integral formalism is employed, and a memoryless and temperature-dependent damping kernel is obtained when the Ohmic spectral density and the regime of very low-temperature are combined. Surprisingly, the real part of the influence functional, i.e. the part related to diffusion, turns out to be temperature independent and non-local in time even in the high-temperature limit, when an Ohmic spectral density is considered. In order to obtain the usual Langevin equation for the particle in the high-temperature regime, a spectral density that is non-vanishing in the limit of zero frequency must be taken.

The modifications implemented in Ref.~\cite{Duarte2006PRL} might appear, at first glance, smoother when compared to those just mentioned. However, they allow for the modeling of an effective coupling between two Brownian particles through their common reservoir. Although this had been predicted long ago from Hydrodynamics \cite{Felderhof1977PhysicaA,Felderhof1978JPA}, it was something missing within the description of Brownian motion through the standard CLM. The essential features introduced in this work are: the nonlinear coupling between the particle and the reservoir degrees of freedom and a spectral density that has both frequency and wave-number dependence. The inspiration for these two ingredients was taken from Ref.~\cite{Hedegard1987PhysScr}, the very first reference mentioned in this Section. The article discusses only the classical regime (part of the quantum effects were investigated later in Refs.~\cite{Duarte2009PRA,Valente2010PRA}) but the obtained results are no less interesting. It is shown that the dynamics of each of the particles is governed by generalized Langevin equations that are coupled through a cross-dissipative term and a potential energy that depends on and decays with the distance between the particles. The authors leave the door open for potential applications in condensed matter physics such as the formation of Cooper pairs, bipolarons etc, where the coupling between particles mediated by their common medium leads to interesting phenomena.

Finally, a relevant variation of the well-known dissipative two-level system or spin-boson model \cite{Leggett1987RMP} was tackled in Ref.~\cite{Brito2008NJP} using a \emph{structured} environment. This model can describe, e.g. a physical situation where the energy scales of the two-level system and a measurement device detecting its state are comparable. Since the device also suffers from the dissipative effects of the environment, the two-level system plus the detector constitutes an efficient channel for decoherence to take place. For example, this mechanism appears in superconducting qubits due to their coupling to the readout dc-SQUID \cite{Chiorescu2003Science}. In this case, the dissipative dynamics of the two-level system can be described as one in which the effective spectral density of the environment presents a pronounced peak at a characteristic frequency. In Ref.~\cite{Brito2008NJP}, the authors present an alternative approximation scheme to describe the dissipative dynamics of the spin-boson model with a structured environment.

\section{Quantum Brownian motion at the origin of quantum thermodynamics}

In the preceding sections, we have seen that quantum Brownian motion has been an active field of research with dedicated contributions spanning over sixty years. While its longevity and appeal certainly originate in fundamental questions about how quantum systems interact with their environment, the impact of quantum Brownian motion in the development of quantum information and quantum thermodynamics can hardly be underestimated. Thus, we conclude this review with a brief summary of the, arguably, most important areas of modern research in which Caldeira's work laid the groundwork and cornerstones.

\subsection{From quantum Brownian motion to understanding Decoherence}

It should not come as a surprise that understanding quantum Brownian motion is essential in the development of so-called fault-tolerant quantum computers. It has been argued that we are currently in the so-called NISQ era, where NISQ stands for ``noisy, intermediate scale quantum'' \cite{Preskill2018quantum}. In many quantum computing platforms, the predominant source of noise arises from the interaction of the quantum system with its environment, which leads to dissipation and decoherence. In particular, mitigating quantum decoherence \cite{Zurek2002RMP,Schlosshauer2005RMP,Gamble2009AJP,Schlosshauer2019PR} poses a significant challenge. In lose terms, ``decoherence'' refers to the decay of quantum superpositions aka quantum coherences of the logical states, which are instrumental in achieving any form of quantum advantage. 

Mathematically, these quantum coherences are described by the off-diagonal matrix elements of the density operator $\rho(t)$. To describe their dynamics, we need to consider the corresponding quantum master equations, which can be written as
\begin{equation}
i\hbar\,\dot{\rho}(t)=\com{H}{\rho(t)}+\mc{D}(\rho(t))\,,
\end{equation}
where $H$ is again the Hamiltonian and $\mc{D}(\rho(t))$ describes the interaction with the environment. Already in Ref.~\cite{Caldeira1983PhysicaA} the Caldeira-Leggett master equation was obtained, which can be written as
\begin{equation}
\label{eq:CL_ME}
i\hbar\,\dot{\rho}(t)=\com{H}{\rho(t)}+\eta\,\com{x}{\{p,\rho(t)\}}-\frac{2 im\,\eta}{\hbar\beta}\,\com{x}{\com{x}{\rho(t)}}\,,
\end{equation}
where we used the same notation as above. 

The Caldeira-Leggett master equation \eqref{eq:CL_ME} is one of the earliest examples to describe the dynamics of a density operator explicitly accounting for environmental noise. Without going too much into detail, the derivation of Eq.~\eqref{eq:CL_ME} requires several approximations that lead to unphysical behavior. For instance, at low temperatures and for early times, Eq.~\eqref{eq:CL_ME} violates the complete positivity required of any valid quantum map \cite{Diosi1993PhysicaA}. To remedy this issue, several different solutions have been proposed, see for instance Refs.~\cite{Diosi1993PhysicaA,Nicacio2024PRE,Ferialdi2017PRA} (we take the opportunity to point out that Caldeira's name is misspelled in the title of Ref.~\cite{Diosi1993PhysicaA}). Interestingly, the quantum master equation can be derived exactly, without approximations or unphysical violations of positivity, if the quantum particle is described by a harmonic oscillator. This exact master equation for the Caldeira-Leggett model has become known as Hu-Paz-Zhang master equation \cite{Hu1992PRD,Halliwell1996PRD}.

In any case, master equations such as Eq.~\eqref{eq:CL_ME} are instrumental in describing the loss of quantum coherence into the environment. Interestingly, Caldeira's own work was often motivated by experimentally realizable scenarios. For instance, Ref.~\cite{CastroNeto1990PRA} analyzed the rapid decay of coherences in the Fock basis for an electromagnetic mode in a cavity. Castro Neto and Caldeira \cite{CastroNeto1990PRA} then also showed that ``higher coherences'', that is coherences farther away from the diagonal of the density operator decay faster. References~\cite{Moussa1996PLA} and \cite{Oliveira2006PRA} focused on different versions of the Stern-Gerlach experiment, and Ref.~\cite{Westfahl2004PRB} on quantum dots.

Given the variety of ways a quantum system can interact with its environment and the importance of understanding decoherence in protecting quantum information, research activities have hardly slowed down. For instance, Ref.~\cite{DavilaRomero1997PRA} analyzed the effect of initial correlations, whereas Ref.~\cite{Halliwell1997PRD} focused on the ``post-decoherence'' dynamics. More recently, research in quantum Brownian motion has, e.g., analyzed the emergence of classical objectivity \cite{Blume2008PRL}, the effects of measurements \cite{BrunoBellomo2007JCF,Ford2007PRA,Magazzu2018NJP}, and the interplay of decoherence and polarization \cite{Sinha2020PRA,Jakubec2025PRA}.

\subsection{A foundational model for Quantum Thermodynamics}

Over the last decade, \emph{Quantum Thermodynamics} \cite{Deffner2019book} established itself as the prevalent theory to describe the thermal properties of quantum systems. Rather naturally, quantum Brownian motion has played an instrumental role in the birth of this timely and active field of research.

The central notion of thermodynamics is entropy, whose continual growth characterizes the passage of time. The natural question arises whether quantum and classical systems differ in their thermodynamic entropy and how it is produced. However, classical thermodynamics is a ultra-weak coupling theory, which means that conventional contributions to the thermodynamic quantities arising from surface effects and interactions are neglected. This is generally no longer a valid approximation for quantum systems, and hence a careful analysis of quantum entropy is necessitated. 

Addressing this issue, Dur\~ao and Caldeira \cite{Durao2016PRE} elucidated how to best define thermodynamic entropy in quantum Brownian motion. In particular, they argued that for dissipative systems a ``system-plus-environment'' method is required from which then (generalized) notions of entropy can be maximized. Interestingly, Dur\~ao and Caldeira \cite{Durao2016PRE} then argue that in the strong coupling regime no generalized entropy permits ``\emph{a proper generalization of thermostatistics}", as different notions might lead to contradicting characterization.

Building upon the von Neumann entropy and taking a more dynamical approach, Weiderpass and Caldeira \cite{Weiderpass2020PRE} then developed a path-integral method to evaluate the entropy production. For the Hu-Paz-Zhang scenario, i.e., for quantum Brownian motion of a single harmonic oscillator coupled to a bosonic bath, they find exact results. The exact expressions then allow to take the systematic high-temperature, classical limit, which gives additional credence to the von Neumann entropy as thermodynamic notion. Complementing this work, Ref.~\cite{Weiderpass2020PRB} studies the heat transport through a harmonic chain. Again leveraging analytical solutions, the quantum-to-classical limit can be explored, which allows to identify the genuinely quantum contributions to the heat current.

In fact, how to properly identify the thermodynamic entropy in quantum Brownian motion is an intricate problem. It was shown that somewhat naive definitions would lead one to conclude that the laws of thermodynamics appear to be broken in the quantum regime \cite{Nieuwenhuizen2002PRE,hanggi2006}. Showing that this is not the case, required significant research efforts \cite{Horhammer2005JPA,Horhammer2008JSP,Pucci2013JSM,Hsiang2018PRE,Colla2021PRA,Qiu2021CTP,Yao2024PRB}, in which Caldeira's contributions \cite{Durao2016PRE,Weiderpass2020PRE} played an important role.

\section{Concluding remarks}

In this brief review, we highlighted Caldeira's numerous contributions to quantum Brownian motion, from foundational work to applications in topical fields like condensed matter physics, quantum information and quantum thermodynamics. Over more than four decades, his work has shaped our understanding of how quantum systems interact with thermal environment, which is the origin of their noisy dynamics. Quite remarkably, Caldeira's research followed a clear and logical pathway starting with the inception of microscopic models that eventually led to the emergence of decoherence theory and quantum thermodynamics. Therefore, the legacy of the Caldeira-Leggett model will continue playing a prominent role in Quantum Brownian motion.

\bmhead{Acknowledgements}

M.V.S.B. ackowledges support from CNPq (Conselho Nacional de Desenvolvimento Cient\'ifico e Tecnol\'ogico) under Grant No. 304120/2022-7. S.D. acknowledges support from the John Templeton Foundation under Grant No. 63626.


\end{document}